\documentclass{article}
\usepackage{multicol,caption}
\usepackage[utf8]{inputenc}
\usepackage{graphicx} 
\usepackage{wrapfig} 
\usepackage[font=footnotesize]{caption}
\usepackage{amssymb,amsmath,amsthm}
\usepackage{enumerate}
\usepackage{anysize}
\usepackage{array}
\usepackage{float}
\usepackage{url}
\usepackage{physics}
\usepackage{mathtools}
\usepackage[usenames,dvipsnames,svgnames,table]{xcolor}
\usepackage[colorlinks=true, allcolors=blue]{hyperref}
\usepackage{geometry}
\usepackage{caption}
\usepackage{subcaption}
\usepackage{fancyhdr}
\usepackage{multirow} 
\usepackage[mathscr]{euscript} 
\newenvironment{Figure}  
{\par\medskip\noindent\minipage{\linewidth}}
{\endminipage\par\medskip}
\title{\Huge Singularity removal in a quantum effective evolution of the Mixmaster cosmological model}
\author{Héctor Hugo Hernández Hernández$^1$, Gustavo Alejandro Sánchez Herrera$^2$  \\ email: 
{$^{1}$hhernandez@uach.mx, $^{2}$cbi2233805002@xanum.uam.mx } 
}
\date{\today}
\begin{document}
\maketitle
\begin{center}
$^{1,2}$\textit{Universidad Autónoma de Chihuahua, Facultad de Ingeniería,} \\ \textit{Nuevo Campus Universitario, Chihuahua 31125, México.}
\end{center}

\begin{abstract}
In this work we analyze the evolution of the quantum Mixmaster cosmological model within an effective approach.  In particular, we study the behaviour of the scale factor and anisotropies of the theory, and determine how it deviates from its classical counterpart due to quantum back-reaction. Remarkably, we determine that the effective evolution avoids the initial singularity. The semiclassical dynamic of the system is obtained from a Hamiltonian in an extended phase space, whose classical position and momentum variables are the expectation values of the corresponding quantum operators, as well as of quantum dispersions and correlations of the system, and is in this framework that we obtain semiclassical one-particle trajectories. 
\end{abstract}
\vspace{2pc}

%
%

\section{Introduction.}

In the mid-20th century Evgeny Lifshitz discovered that as the universe shrinks to regions of space close to the initial sigularity, spacetime is no longer isotropic \cite{kiefer2018dynamics}. Because of this, the interest in anisotropic cosmological models increased considerable during the past years \cite{koshelev2018towards, toporensky2002nature, saha2006anisotropic, himmetoglu2009instability}. One of the most prominent results of these investigations is the BKL conjecture, stating that, in a general way, it is possible to neglect the matter terms near the initial singularity because, for them, time derivatives are dominant over those with spatial derivatives; the dynamics in this case is described by the Bianchi IX model \cite{ringstrom2001bianchi}. In recent years, a great variety of studies of anisotropic models of the universe, based on this conjecture, have been carried out \cite{piechockiquantum, ritchie2022bianchi, alonso2021quantum}. 

The most general anisotropic homogeneous model of the universe, based on the BKL conjecture, is the Mixmaster \cite{giovannetti2019polymer}. It describes the behavior of the universe near the initial singularity. The universe is treated as a point particle moving through the anisotropies space, subject to a time-dependent potential \cite{misner1994mixmaster}.
The studies carried out on the classic Mixmaster model have shown that it is not only singular, but also has a chaotic behavior close to the initial singularity \cite{cornish1997mixmaster, barrow1982chaotic, zardecki1983modeling}, therefore, a quantization scheme is introduced in the anticipation of mitigating these issues. Among the most interesting approaches in the analysis of these quantum cosmological models are effective quantization schemes based on loop quantum gravity \cite{brunnemann2006cosmological}, and effective polymeric quantum mechanics \cite{lecian2013semiclassical}, showing that the initial singularity is removed, and the chaos present in the theory is reduced \cite{bojowald2023chaotic} This shows the importance of effective quantization approaches in analyzing complex cosmological models, as the ones mentioned above.

Quantum effective methods in quantum mechanics allow us to obtain an approximate solution of the whole system. In particular, momenta quantum mechanics reduces quantum systems to semiclassical ones, where the dynamics is obtained from an effective Hamiltonian in an extended phase space \cite{bojowald2006effective}. One of the most prominent features of this method is that the notion of individual particle-trajectories is recovered, a characteristic not existing in usual quantum mechanics. This trajectories describe the evolution of expected values of position and momentum operators, and of the (infinite many) quantum dispersions. The versatility of application of this method has allowed the study of a broad spectrum of quantum systems, ranging from the relatively simple phenomenon of quantum tunneling \cite{aragon2020effective}, to models of quantum cosmology \cite{bojowald2011high, brizuela2019moment, brizuela2022semiclassical}.

In this work we obtain a system of effective equations of motion for spatial anisotropies and the scale factor of the Mixmaster model, once we determine the effective extended Hamiltonian. The interaction of the several degrees of freedom of the system is encoded in an effective potential, obtained in a direct way. In section 2 we review the more general aspects of the classical Mixmaster model and perform a canonical transformation to explicitly express the Hamiltonian of the theory as a kinetic plus potential term. In section 3 we provide the formalism of momenta quantum mechanics, and we apply this to the Mixmaster model in order to obtain the effective dynamics of the system. Finally, in section 4 the semiclassical evolution is analyzed.

\section{Classical Mixmaster model}\label{seccion2}

The interest in investigating the Mixmaster model lies in its representation as a general solution to Einstein's equations near the initial singularity \cite{battisti2009mixmaster}. Furthermore, various studies have been conducted that provide substantial evidence supporting the BKL conjecture \cite{guo2014spherical, heinzle2012spike, ashtekar2011hamiltonian}. We will analyze the quantum model with an effective prescription of quantum mechanics, so we start with a discussion of the classical model.
	
The most general anisotropic cosmological model is the Bianchi $IX$, whose metric is given by \cite{misner1994mixmaster}
\begin{eqnarray}\label{BianchiIXmetrics}
	ds^{2}&=& -N^{2}(t)dt^{2} + g_{ij}(t)\sigma_{i}\sigma_{j},
\end{eqnarray} 
where $N$ is the lapse function, and $\sigma$ are differential forms of the three-sphere \cite{misner1969mixmaster}
\begin{eqnarray}
	\sigma_{1}&=&  \cos(\psi) d\theta+\sin(\psi)\sin(\theta)d\phi,   \nonumber \\
	\sigma_{2}&=&  \sin(\psi) d\theta-\cos(\psi)\sin(\theta)d\phi,   \nonumber \\
	\sigma_{3}&=& d\psi + \cos(\theta) d\phi   .		 \nonumber
\end{eqnarray}
In particular, the metric $g_{ij}$ can be written in terms of the Misner parameters $\alpha$, and $\beta$ such that (\ref{BianchiIXmetrics}) is 
\begin{eqnarray}\label{BianchiIXMetric}
	ds^{2}= -N^{2}dt^{2}+ e^{2 \alpha}( e^{2\beta})_{i j}\sigma^{i}\sigma^{j}, 
\end{eqnarray}
where $\alpha$ is the parameter determining the volume of the universe, and $\beta_{ij}$ is a null trace matrix $\tr(\beta_{ij})=0$ containing the spatial anisotropies. The components of the matrix $\beta_{ij}$ satisfy  the following equations \cite{moriconidynamical} 
\begin{eqnarray}\label{misnerv}
	\beta_{11}&=& \beta_{+} + \sqrt{3} \beta_{-},  \nonumber \\
	\beta_{22}&=& \beta_{+} - \sqrt{3} \beta_{-}, \nonumber \\
	\beta_{33}&=& -2\beta_{+}.
\end{eqnarray}
On the other hand, the square root of the determinant of the metric $\sqrt{\det(g_{ij})}$ from eq. (\ref{BianchiIXMetric}) allow us to identify the volume of the universe as the scale factor $a(t)$ in terms of the parameter $\alpha$ \cite{misner1994mixmaster} 
\begin{eqnarray}\label{atexp3}
	a(t) = \sqrt{\det(g_{ij})} = \exp(3 \alpha).
\end{eqnarray}
The relation between the parameters $\alpha, \beta_{+}, \beta_{-}$ with the scale factor $a(t)$ is \cite{moriconidynamical}
\begin{eqnarray}\label{UniverseRatioParametrization}
	\alpha(t) &=& \frac{1}{3} \ln(a(t)), \nonumber  \\
	\beta_{+}(t)	&=&  \frac{1-P_{3}}{2} \alpha(t)   	, \nonumber \\
	\beta_{-}(t)	&=&	 \sqrt{3}(P_{1}-P_{2}) \alpha(t)		,
\end{eqnarray} 
where $P_{i}$ are the Kasner coefficients. For instance, to describe a universe that expands in two directions, and contracts in  the other, the coefficients $P_{i}$ are $P_{1}=P_{3} = 2/3$, and $P_{2}=-1/3$ \cite{thorne2000gravitation}. The equation (\ref{atexp3}) shows that as $\alpha \rightarrow-\infty$, $a(t)\rightarrow 0$, which corresponds to the initial singularity 

When the Hamiltonian formulation for the Bianchi IX model is introduced, the resulting model is the Mixmaster. Its dynamics is determined by a Hamiltonian $\mathcal{H}$, which is obtained through the variation of the following action \cite{moriconidynamical}
\begin{eqnarray}\label{MixmasterAction}
	I&=& \int dt \left(p_{\alpha}\dot{\alpha} + p_{+}d\dot{\beta_{+}} +p_{-}d\dot{\beta_{-}}- N \mathcal{H}   \right), \nonumber \\
\end{eqnarray}
where $p_{\alpha}$, and $p_{\pm}$ are the conjugate momenta of $\alpha$ and $\beta_{\pm}$ \cite{brizuela2022semiclassical}. $\mathcal{H}$ is defined as  
\begin{eqnarray}\label{HIx}
	\mathcal{H} &=& \frac{k}{3(8\pi)^{2}}e^{-3\alpha} \left(-p_{\alpha}^{2} + p_{+}^{2}+p_{-}^{2}+ \mathcal{V} \right), 
\end{eqnarray}
with
\begin{eqnarray}\label{bianchispotentials}
	\mathcal{V} &=& \frac{3(4\pi)^{2}}{k^{2}} e^{4\alpha} V(\beta_{\pm}), 
\end{eqnarray}
where $k = 8 \pi G$ and $V$ is the potential. Its particular form for  the Bianchi $IX$ model \cite{ryan2015homogeneous} is the following  
\begin{eqnarray}\label{PvT}
	V(\beta_{\pm}) &=& e^{-8\beta_{+}}- 4e^{-2\beta_{+}}\cosh(2\sqrt{3}\beta_{-}) 
	+ 2e^{4 \beta_{+}}\left[\cosh(4\sqrt{3}\beta_{-})-1\right].
\end{eqnarray}
This potential is a function of anisotropies $\beta_{+}, \beta_{-}$, and is graphically represented by equilateral triangles that evolve in time as shown in the figure \ref{IXPotentialss}. In general, $\mathcal{H}$ contains the physics of all the Bianchi universes in the potential term $\mathcal{V}$ from the equation (\ref{bianchispotentials}), where to each Bianchi model corresponds a different equipotential line \cite{battisti2009mixmaster}. 
\begin{Figure}
	\centering
	\captionsetup{type=figure}
\includegraphics[width=0.45\linewidth]{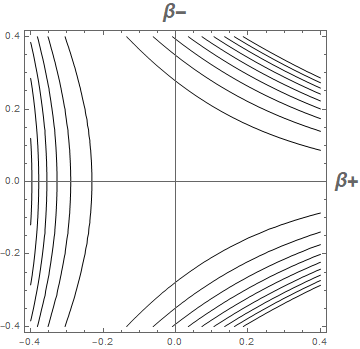}
	\captionof{figure}{Parametric representation of the Mixmaster potential. The equipotential lines shows how the potential evolve in time.}
  \label{IXPotentialss}
 \end{Figure}

The dynamics is obtained through the variation of ($\ref{MixmasterAction}$) with respect to each of the variables. The variation with respect to the lapse function $N$ generates the Hamiltonian constraint $\mathcal{H}= 0$. Solving for $p_{\alpha}$ we have
\begin{eqnarray}\label{parestriction}
	-p_{\alpha}= H_{IX} = \sqrt{p_{+}^{2}+p_{-}^{2}+\mathcal{V}}.
\end{eqnarray}
From this expression one obtains the dynamics of the universe represented as a point particle subject to a variable potential in the space of anisotropies $\beta_{+}$, and $\beta_{-}$. Each of the classical variables evolves with respect to the parameter $\alpha$. Close to the initial singularity, it is possible to show that the particle moves twice as fast as the separation movement of the  potential walls, and then it experiences consecutive reflections as the volume of the universe decreases \cite{moriconidynamical}. 

\begin{figure}[ht]
        \centering  
   	\begin{subfigure}[b]{0.47\textwidth}
   		\centering   		\includegraphics[width=\textwidth]{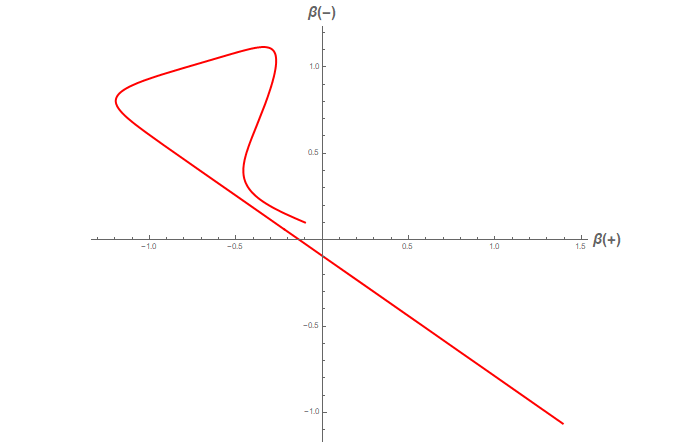}
   		\caption{The initial conditions are $\beta_{+}=-0.1$, $\beta_{-}=0.1$, $p_{\beta_{+}}=-3, p_{\beta_{-}}= 1.28$. The value of $V(\beta_{+},\beta_{-})$ is of the order of $\exp{-4 \alpha}$}
   		\label{1trayectoryClassical}
   	\end{subfigure}
   	\hfill
   	\begin{subfigure}[b]{0.47\textwidth}
   		\centering
   \includegraphics[width=\textwidth]{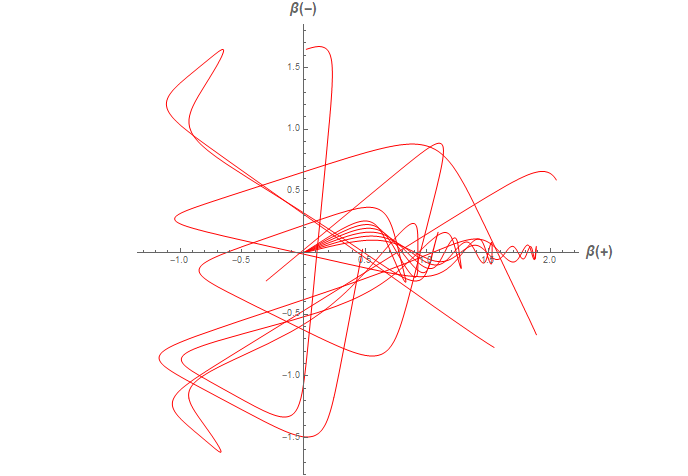}
   		\caption{The initial conditions are $\beta_{+}=\beta_{-}=0$, $p_{\beta_{+}}=100, p_{\beta_{-}}=8$. The value of $V(\beta_{+},\beta_{-})$ is of the order of $\exp{-1.2 \alpha}$. }
   \label{ManyTrayectorieClassica}
   	\end{subfigure}
   	\caption{Trajectory of the particle in space $(\beta_{+}, \beta_{-})$ for different values of $V(\beta_{+},\beta_{-})$. It is shown how the temporal dependence of $V(\beta_{+},\beta_{-})$ results in reflection in different equipotential lines.}
   	\label{Classicaltrayectories}
   \end{figure}

Given the Hamiltonian in (\ref{parestriction}), and using the potential $V(\beta_{\pm})$ from ($\ref{PvT}$), we get the following set of classical equations of motion for the Mixmaster model  
\begin{eqnarray}\label{motioneqHbIX}
	\dot \beta_{+}&=&  \frac{p_{+}}{H_{IX}}    ,\nonumber  \\
	\dot \beta_{-}&=& \frac{p_{-}}{H_{IX}}   ,\nonumber  \\
	\dot p_{+}&=& 4\frac{3(4\pi)^{2}}{k^{2}H_{IX}} e^{4\alpha} \left[e^{-8\beta_{+}}-e^{-2\beta_{+}}\cosh(2\sqrt{3}\beta_{-}) \right. \nonumber \\  &-&e^{4\beta_{+}} \left. \left(\cosh(4\sqrt{3})-1\right) \right]   ,\nonumber  \\
	\dot p_{-}&=& 4\sqrt{3}\frac{3(4\pi)^{2}}{k^{2}H_{IX}} e^{4\alpha} \left[e^{-2\beta_{+}}\sinh(2\sqrt{3}\beta_{-}) \right. \nonumber \\ &-& e^{4\beta_{+}}\left. \sinh(4\sqrt{3}) \right]     .
\end{eqnarray}
This is a highly non-linear, coupled system for the classical anisotropies and their momenta, so we solve it numerically. The evolution obtained is shown in figure \ref{Classicaltrayectories}. Figure \ref{1trayectoryClassical} shows that the trajectory of the universe experience reflections at the potential barriers, while in the figure \ref{ManyTrayectorieClassica} we can see the chaotic behaviour of the classical system for different given trajectories. As the universe approaches the initial singularity, the reflections of the particle with the potential barriers decrease, and the trajectory of the universe behaves as a straight line in space of anisotropies \cite{bojowald2023chaotic}, while the parameter $\alpha $ tends to minus infinity. The initial singularity corresponds to $\abs{\beta_{\pm}} \rightarrow \infty $. 

\begin{figure}[ht]
   	\centering
   	\begin{subfigure}[b]{0.28\textwidth}
   		\centering
   	\includegraphics[width=\textwidth]{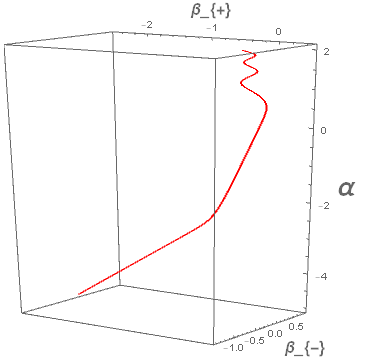}
   		\caption{ The initial conditions are $x(0)= 0.1$, $y(0) = 0.1$, $px(0) = -60$, $py(0) = 200$.}
   		\label{3D1}
   	\end{subfigure}
   	\hfill
   	\begin{subfigure}[b]{0.28\textwidth}
   		\centering   		\includegraphics[width=\textwidth]{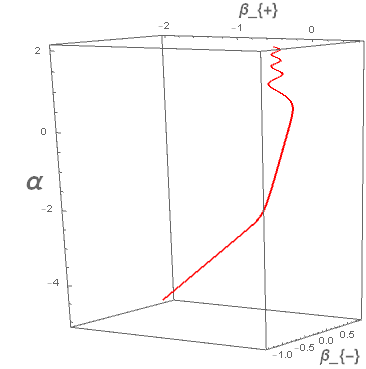}
   		\caption{ The initial conditions are $x(0) = 0.1$, $y(0) = 0.1$, $px(0) = -80$, $py(0) = 200$.}
   		\label{3D2}
   	\end{subfigure}
    \hfill
   	\begin{subfigure}[b]{0.28\textwidth}
   		\centering
   \includegraphics[width=\textwidth]{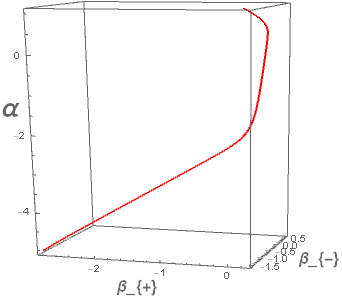}
   		\caption{ The initial conditions are $x(0) = 0.1$, $y(0) = 0.1$, $px(0) = -30$, $py(0) = 200$.}
   		\label{3D3}
   	\end{subfigure}
   	\caption{Trajectory of the particle in space $(\beta_{+},\beta_{-}, -\alpha)$. The pictures shows that for any initial condition, as time goes backwards, $\alpha$ goes to $-\infty$, and the classical singularity is reached.}
   	\label{C3D}
   \end{figure}

   In figure \ref{C3D} those trajectories are plotted in three-dimensional space $(\beta_{+}, \beta_{-}, -\alpha)$.
   
Applying the canonical transformations $q=\exp{(3/2)\alpha}$, $p=\frac{3}{2}\exp{-(3/2)\alpha}p_{\alpha}$ for the variables $\alpha$, and $p_{\alpha}$, and ${p'}_{+}=p_{+}/q$, $\beta_{+}^{'}=\beta_{+}q$, ${p'}_{-}=p_{-}/q$, $\beta_{-}^{'}=\beta_{-}q$ for the anisotropies and their momenta,
the Hamiltonian constraint (\ref{HIx}) reduces to the following \cite{de2023mixmaster}
\begin{eqnarray}\label{TrHam}
	\mathcal{K} &=& -\frac{9}{4}p^{2} + {p'}_{+}^{2}+{p'}_{-}^{2}+ \frac{3(4 \pi)^{4}}{k} q^{2/3}V(q,{\beta'}_{+},{\beta'}_{-}), 
\end{eqnarray}
where the potential $V({q,\beta'}_{+},{\beta'}_{-})$ is given by
\begin{eqnarray}\label{TpOT}
	V(q,{\beta'}_{+},{\beta'}_{-}) &=& e^{-\frac{8{\beta'}_{+}}{q}}- 4e^{-\frac{2{\beta'}_{+}}{q}}\cosh(\frac{2\sqrt{3}{\beta'}_{-}}{q}) 
	+ 2e^{\frac{4{\beta'}_{+}}{q}}\left[\cosh(\frac{4\sqrt{3}{\beta'}_{-}}{q})-1\right].
\end{eqnarray}

In the isotropic limit  ${\beta'}_{+}={\beta'}_{-}={p'}_{+}={p'}_{-}=0$, the potential $V(\beta_{+},\beta_{-},q)$ is a negative constant with value $V(\beta_{+},\beta_{-},q)=-3$. Then (\ref{TrHam}) becomes
\begin{eqnarray}\label{THIso}
	\mathcal{K}_{iso} &=& \frac{1}{4}p^{2} + 32 \pi^{3} q^{2/3}. 
\end{eqnarray}

When we consider the isotropic part of the system, (\ref{TrHam}) is reduces to (\ref{THIso}), which has only one degree of freedom. From this, the classical equations of motion for $q$ and $p$ are
\begin{eqnarray}\label{TransformedClassical}
	\dot q&=&  \frac{p}{2}    ,\nonumber  \\
	\dot p&=& -\frac{64}{3} \pi^{3}q^{-1/3}. 
\end{eqnarray}
Since $q=\exp{(3/2)\alpha}$, from (\ref{atexp3}) we know that $q=\sqrt{a}$. Therefore, $q \rightarrow 0$ implies that $a \rightarrow 0$ and $Log(q) \rightarrow -\infty$, that corresponds to the initial singularity. In figure \ref{CTK} we show the relation between $Log(q)$ and $p$.
\begin{Figure}
	\centering
	\captionsetup{type=figure}
\includegraphics[width=0.55\linewidth]{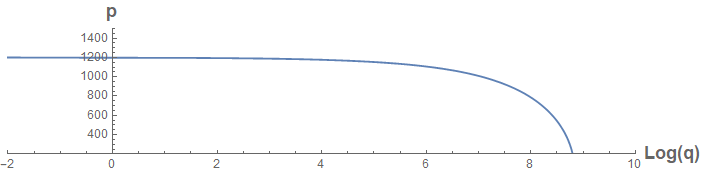}
	\captionof{figure}{Plot between $q$ and $p$. Since $q = \exp{(3/2) \alpha}$, the initial singularity occurs when the $Log(q) \rightarrow -\infty$. The initial conditions are $q(0) = 7000$, $p(0) = 10$.}
  \label{CTK}
 \end{Figure}

\begin{figure}[ht]
   	\centering
   	\begin{subfigure}[b]{0.3\textwidth}
   		\centering
\includegraphics[width=\textwidth]{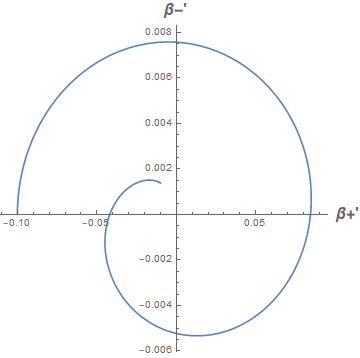}
   \caption{$q(0)=3000$, and $p(0)=10$, ${\beta'}_{+} =-0.1$, ${\beta'}_{-}=0$, ${p'}_{+}=0$, ${p'}_{-}=0.01$.}
   		\label{Cani1}
   	\end{subfigure}
   	\hfill
   	\begin{subfigure}[b]{0.3\textwidth}
   		\centering   		\includegraphics[width=\textwidth]{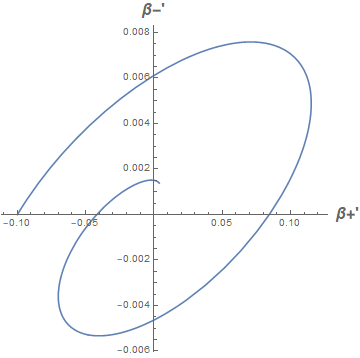}
   		\caption{$q(0)=3000$, and $p(0)=10$, ${\beta'}_{+} =-0.1$, ${\beta'}_{-}=0$, ${p'}_{+}=0.1$, ${p'}_{-}=0.01$.}
   		\label{Cani2}
   	\end{subfigure}
    \hfill
   	\begin{subfigure}[b]{0.3\textwidth}
   		\centering   \includegraphics[width=\textwidth]{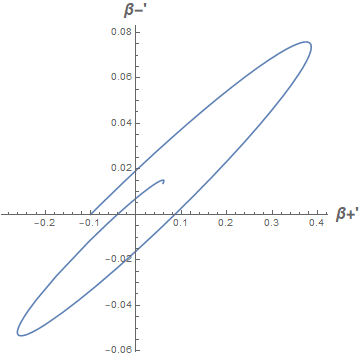}
   		\caption{$q(0)=3000$, and $p(0)=10$, ${\beta'}_{+} =-0.1$, ${\beta'}_{-}=0$, ${p'}_{+}=0.5$, ${p'}_{-}=0.01$.}
   		\label{Cani3}
   	\end{subfigure}
   	\caption{Classical evolution in the space (${\beta'}_{+}$, ${\beta'}_{-}$).}
   	\label{Cani}
   \end{figure}

The equations of motion obtained from the (\ref{TrHam}) are 
 \begin{eqnarray}\label{CmCh}
	\dot q&=&  -\frac{9}{2}p    ,\nonumber  \\
    \dot {\beta'}_{+}&=&  2{p'}_{+}    ,\nonumber  \\
    \dot {\beta'}_{-}&=&  2{p'}_{-}    ,\nonumber  \\
	\dot p&=& -\frac{3(4 \pi)^{4}}{k}q^{-1/3} \left( q\frac{\partial V(q,{\beta'}_{+},{\beta'}_{-})}{\partial q}-\frac{2}{3} \right), \nonumber \\
    \dot {p}'_{+}&=&  -\frac{3(4 \pi)^{4}}{k}q^{2/3} \frac{\partial V({q,\beta'}_{+},{\beta'}_{-})}{\partial {\beta'}_{+}},\nonumber  \\
	\dot {p}'_{-}&=&  -\frac{3(4 \pi)^{4}}{k}q^{2/3} \frac{\partial V({q,\beta'}_{+},{\beta'}_{-})}{\partial {\beta'}_{-}}. 
\end{eqnarray}
In figure \ref{AnClSi} and (\ref{Cani}) we show the classical evolution obtained from (\ref{CmCh}). 
\begin{Figure}
	\centering
	\captionsetup{type=figure}
\includegraphics[width=0.55\linewidth]{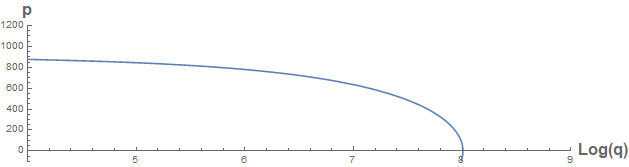}
	\captionof{figure}{$Log(q) $ vs $p$ plot. The initial conditions are $q(0) = 3000$, $p(0) = 10$, ${\beta'}_{+} =-0.1$, ${\beta'}_{-}=0$, ${p'}_{+}=0.1$, ${p'}_{-}=0.1$. }
      \label{AnClSi}
 \end{Figure}

\section{Effective dynamics of the Mixmaster model}

\subsection{Effective momenta quantum mechanics}

In usual quantum mechanics, the Schrödinger equation $\hat H \psi= E \psi  $ determines the evolution of the system: all the information is encoded in the wave function, and classical variables are now operators. Within this Schrödinger representation, the concept of a single particle's position is absent, and also trajectories. It is possible, however, with a generalization of the Ehrenfest theorem, to obtain an effective description by means of a Hamiltonian $H_{Q}=\left\langle \hat H \right\rangle$, which depends on expectation values of observables and quantum dispersions (or momenta) \cite{bojowald2011quantum}. Once this Hamiltonian is obtained, dynamical equations of motion can be obtained in the usual way through the following equation 
\begin{eqnarray}\label{OperatorsBracketComutation}
	\lbrace \langle \hat f \rangle, \langle \hat g \rangle \rbrace&=& \frac{1}{i \hbar}\langle [\hat f, \hat g] \rangle. 
\end{eqnarray}  
For one degree of freedom, these momenta are defined as
\begin{eqnarray}\label{G1par}
	G^{a,b} &:=& \langle  (\hat p-\langle \hat p\rangle )^{a}(\hat q-\langle \hat q\rangle)^{b} \rangle_{\text{Weyl}},
\end{eqnarray}  
where $\langle \hat q\rangle$ and $\langle \hat p\rangle$ are the expectation values of position and momentum respectively, and $a+b \geq 2$. The legend $\text{Weyl}$ means totally symmetrization \cite{bojowald2011high}. 
The momenta $G^{a,b}$ obey a generalization of the Heisenberg uncertainty principle \cite{bojowald2006effective}, that is
\begin{eqnarray}\label{MOmentsUncertanityRelation}
	G^{2,0}G^{0,2} - (G^{1,1})^{2} &\geq& \frac{\hbar^{2}}{4}.
\end{eqnarray}

The effective Hamiltonian $H_{Q}$ is defined in the following way
\begin{eqnarray}\label{H1PAR}
	H_{Q}  &= &  \sum_{a=0}^{\infty} 	\sum_{b=0}^{\infty} \frac{1}{a!b!} \frac{\partial^{a+b}H }{\partial 	p^{a}\partial q^{b}} G^{a,b}.
\end{eqnarray}
A general expression for $k$ degrees of freedom for $H_{Q} $ is  \cite{hernandez2021semiclassical}

\begin{eqnarray}\label{generalHq}
H_{Q}= \sum_{a_{1},b_{1}}^{\infty}\cdots \sum_{a_{k},b_{k}}^{\infty} \frac{1}{a_{1}!b_{1}!\cdots a_{k}!b_{k}!}\frac{\partial^{a_{1}+b_{1}+\cdots +a_{k}+b_{k}}H}{\partial q_{1}^{a_{1}} \partial p_{1}^{b_{1}} \cdots \partial q_{k}^{a_{k}}  \partial p_{k  }^{b_{k}} } G^{a_{1},b_{1},\cdots ,a_{k},b_{k}}.
\end{eqnarray}

Equations of motion are obtained from this effective Hamiltonian. For general systems there are an infinite number of momenta and, correspondingly, an infinite number of equations of motion, usually impossible to solve analytically, although consistent truncations can be implemented in order to obtain approximated solutions \cite{bojowald2012quantum}. 

Analysis of this effective dynamics is done usually in a numerical way, for which initial conditions are required. We employ general squeezed states to determine such conditions, for example, for a Gaussian function $\psi_{\sigma} $ 
\begin{eqnarray}\label{GaussP}
	\psi_{\sigma}(q)&=& \frac{1}{\pi^{1/4} \sqrt{\sigma}} 	\exp{-\frac{(q-\langle q\rangle )^{2}}{2 \sigma^2}+ \frac{i  q \langle  p\rangle }{\hbar}},
\end{eqnarray}
we can readily obtain the initial values for quantum variables. For instance
\begin{eqnarray}\label{GGC}
   G^{a,b}=\left\{
  \begin{array}{@{}ll@{}}
    2^{-(a+b)} \hbar^{a} \sigma^{b-a} \frac{a ! b !}{(a/2)!(b/2)! }, & \text{$a$ and $b$ even} \\
    0, & \text{otherwise}
  \end{array}\right.
\end{eqnarray} 
One can see that the initial momenta $G^{a,b}$ saturate the Heisenberg uncertainty relation
\begin{eqnarray}\label{MomnentUncertSatured}
	G^{2,0}G^{0,2} &=& \frac{\hbar^{2}}{4}. \nonumber
\end{eqnarray}

In this effective formulation, the momenta scale as powers of $\hbar$, that is
\begin{eqnarray}
	G^{a,b} &\propto& \hbar^{\frac{(a+b)}{2}},  \nonumber
\end{eqnarray}

It is possible to obtain important modifications of the classical system considering only the second order terms \cite{bojowald2012quantum}.

On the other hand, the momenta $G^{a,b}$ form a set of non-canonical coordinates in the quantum phase state which complicates the canonical analysis of the system \cite{ding2022effective}. However, it is possible to generalize this effective method through a coordinate transform that allow us to rewrite the momenta in terms of pairs of canonical variables $s_{i}$, $p_{s_{i}}$ and  Casimir parameters \cite{baytacs2020faithful}. These new coordinates encode  all the quantum information of the system. The main advantage of this reformulation is that it avoids the truncation required by the momenta approach, and allows us to construct an effective potential  \cite{baytacs2019effective}
\begin{equation}\label{allorderpotential}
	V_{\text{All}}(q,s) = \frac{1}{8 s^{2}} + \frac{1}{2} 	[V(q+s)+ V(q-s)],
\end{equation}
where $V(q)$ is the classical potential of the system. In general, for 3 degrees of freedom we have \cite{baytacs2018canonical}
\begin{eqnarray}\label{vall2pairs}
	V_{\text{All}}(x_{i},s_{j})&=& \sum_{i=1}^{3} \frac{U}{2s_{i}^{2}} + \frac{1}{8}\left[V(x_{i}+s_{i})+ V(x_{i}-s_{i}) \right].
\end{eqnarray}
For a second order truncation in the momenta $G^{a,b}$ the coordinate transformation is
\begin{eqnarray}
    s &=& \sqrt{G^{0,2}} \nonumber \\
    p_{s}&=& \frac{G^{1,1}}{\sqrt{G^{0,2}}}\nonumber \\
    U&=& G^{0,2}G^{2,0} -{(G^{1,1})}^{2}.
\end{eqnarray}
where $\lbrace s,p_{s}\rbrace =1$, and $\lbrace s,U\rbrace =\lbrace p_{s},U\rbrace =0$ \cite{ding2022effective}.

\subsection{Momenta effective dynamics}

In order to analyze the effective evolution of anisotropies $\beta_{+}$, $\beta_{-}$, we use the Hamiltonian (\ref{parestriction})
\begin{eqnarray}\label{HIXxy}
	H_{IX} = \sqrt{p_{x}^{2}+p_{y}^{2}+\frac{3(4\pi)^{2}}{k^{2}} e^{4\alpha} V(x,y)},
\end{eqnarray}
where the potential $V(x,y)$ is
\begin{eqnarray}
	V(x,y) &=& e^{-8x}- 4e^{-2x}\cosh(2\sqrt{3}y) 
	+ 2e^{4x}\left[\cosh(4\sqrt{3}y)-1\right].
\end{eqnarray}
We have rewritten (for simplicity and to facilitate the numerical application of the method) $\beta_{+} \rightarrow x$, $\beta_{-} \rightarrow y, p_{+} \rightarrow p_{x}$, and $p_{-} \rightarrow p_{y}$ . 

Using the Hamiltonian (\ref{HIXxy}), and truncating to second order in momenta the equation (\ref{generalHq}), we obtain the following expression

\begin{eqnarray}\label{eHM}
	H_{QIX} &=& H_{IX}+\left(H_{IX}^{-1}-p_{x}^{2}H_{IX}^{-3}\right)G^{0200}+\left(H_{IX}^{-1}-p_{y}^{2}H_{IX}^{-3}\right)G^{0002} + \eta(\alpha) \left[2\frac{\partial}{\partial x} \left(g_{1}(x,y)H_{IX}^{-1}   \right)G^{2000} \right. \nonumber \\
	&+&  2\sqrt{3}\frac{\partial}{\partial y} \left(g_{2}(x,y)H_{IX}^{-1}   \right)G^{0020} - \left. \frac{4p_{x}}{H_{IX}^{3}} g_{1}(x,y) G^{1100} - \frac{4\sqrt{3}p_{y}}{H_{IX}^{3}} g_{2}(x,y) G^{0011} \right].  
\end{eqnarray}
The functions $g_{1}(x,y)$, $g_{2}(x,y)$, and $\eta(\alpha)$ are as follows
\begin{eqnarray}
	g_{1}(x,y) &=& -e^{-8x}+e^{-2x}\cosh(2\sqrt{3}y)+e^{4x}\left(\cosh(4\sqrt{3}y)-1\right) \nonumber \\ 
	g_{2}(x,y) &=& -e^{-2x}\sinh(2\sqrt{3}y)+e^{4x}\sinh(4\sqrt{3}y),\nonumber \\
	\eta(\alpha) &=&  \frac{3(4\pi)^{2}}{k^{2}} e^{4\alpha},  
\end{eqnarray}
and the momenta for two degrees of freedom is given by
\begin{eqnarray}\label{G2par}
	G^{a b c d} &:=& \langle  (\hat p_{x}-\langle \hat p_{x}\rangle )^{a}(\hat x-\langle \hat x\rangle)^{b} (\hat p_{y}-\langle \hat p_{y}\rangle )^{c}(\hat y-\langle \hat y\rangle)^{d} \rangle_{\text{Weyl}}.
\end{eqnarray}

The equations of motion for the classical variables are 

\begin{eqnarray}\label{EFEDynamicsiX}
	\dot{x} &=& \frac{p_{x}}{H_{IX}}-2p_{x}\eta(\alpha)\frac{\partial}{\partial x
	}\left(g_{1}(x,y)H_{IX}^{-3}\right)  G^{2000} - 4\eta(\alpha)g_{1}(x,y)\frac{\partial}{\partial p_{x}}\left(p_{x}H_{IX}^{-3}\right) G^{1100}\nonumber \\
	&+& \frac{1}{2}\frac{\partial}{\partial p_{x}}\left(H_{IX}^{-1}-p_{x}^{2}H_{IX}^{-3}\right)G^{0200}  +  2\sqrt{3}\eta(\alpha)\frac{\partial^{2}}{\partial p_{x} \partial_{y}}\left(g_{2}(x,y)H_{IX}^{-1}\right) G^{0020}\nonumber \\
	&-& 4\sqrt{3} p_{y}\eta(\alpha) \frac{\partial}{\partial p_{x}}\left(H_{IX}^{-3}\right) g_{2}(x,y) G^{0011} + \frac{1}{2}\frac{\partial}{\partial p_{x}}\left(H_{IX}^{-1}-p_{y}^{2}H_{IX}^{-3}\right) G^{0002}  ,\nonumber  \\
	\dot{y} &=& \frac{p_{y}}{H_{IX}} + 2\eta(\alpha)\frac{\partial^{2}}{\partial x \partial p_{y}
	}\left(g_{1}(x,y)H_{IX}^{-1}\right) G^{2000} - 4p_{x}\eta(\alpha)\frac{\partial}{\partial p_{y}}\left(g_{1}(x,y)H_{IX}^{-3}\right)G^{1100} \nonumber \\
	&+& \frac{1}{2}\frac{\partial}{\partial p_{y}}\left(H_{IX}^{-1}-p_{x}^{2}H_{IX}^{-3}\right) G^{0200} + 2\sqrt{3}\eta(\alpha)\frac{\partial^{2}}{\partial p_{x} \partial_{y}}\left(g_{2}(x,y)H_{IX}^{-1}\right) G^{0020} \nonumber \\
	&+& \frac{\partial}{\partial y}\left(H_{IX}^{-1}-p_{y}^{2}H_{IX}^{-3}\right) G^{0011} +\frac{1}{2}\frac{\partial}{\partial p_{y}}\left(H_{IX}^{-1}-p_{y}^{2}H_{IX}^{-3}\right)G^{0002} ,\nonumber  \\ 
	\dot{p_{x}} &=& -4\eta(\alpha)g_{1}(x,y)H_{IX}^{-1} - 2 \eta(\alpha) \frac{\partial^{2}}{\partial x^{2}}\left(g_{1}(x,y)H_{IX}^{-1}\right)  G^{2000} - 4 \eta(\alpha) \frac{\partial^{2}}{\partial x \partial p_{x}}\left(g_{1}(x,y)H_{IX}^{-1}\right)G^{1100} \nonumber \\
	&-& \frac{1}{2}\frac{\partial}{\partial x}\left(H_{IX}^{-1}-p_{x}^{2}H_{IX}^{-3}\right)G^{0200}-2\sqrt{3}\eta(\alpha)\frac{\partial^{2}}{\partial x \partial y}\left(g_{2}(x,y)H_{IX}^{-1}\right)G^{0020} \nonumber \\
	&+& 4\sqrt{3} p_{y} \eta(\alpha) \frac{\partial}{\partial x }\left(g_{2}(x,y)H_{IX}^{-3}\right) G^{0011} -\frac{1}{2}\frac{\partial}{\partial x}\left(H_{IX}^{-1}-p_{y}^{2}H_{IX}^{-3}\right)G^{0002} ,\nonumber  \\
	\dot{p_{y}} &=& -4\sqrt{3}\eta(\alpha)g_{2}(x,y)H_{IX}^{-1} - 2 \eta(\alpha) \frac{\partial^{2}}{\partial x \partial y}\left(g_{1}(x,y)H_{IX}^{-1}\right)G^{2000} +4 p_{x} \eta(\alpha) \frac{\partial}{\partial y }\left(g_{1}(x,y)H_{IX}^{-3}\right)G^{1100} \nonumber \\
	&-& \frac{1}{2}\frac{\partial}{\partial y}\left(H_{IX}^{-1}-p_{x}^{2}H_{IX}^{-3}\right)G^{0200} -2\sqrt{3}\eta(\alpha)\frac{\partial^{3}}{\partial y^{3}}\left(g_{2}(x,y)H_{IX}^{-1}\right)G^{0020} \nonumber \\
	&-& 4\sqrt{3}\eta(\alpha)\frac{\partial^{2}}{\partial_{y} \partial p_{y}}\left(g_{2}(x,y)H_{IX}^{-1}\right)   G^{0011} -\frac{1}{2}\frac{\partial}{\partial y}\left(H_{IX}^{-1}-p_{y}^{2}H_{IX}^{-3}\right)G^{0002} ,
\end{eqnarray}
and for the momenta,
\begin{eqnarray}\label{appendix1}
	\dot G^{2000}&=& \frac{8p_{x}}{H_{IX}^{3}}\eta(\alpha)g_{1}(x,y) G^{2000}-2\left(H_{IX}^{-1}-p_{x}^{2}H_{IX}^{-3}\right)G^{1100}  \nonumber \\
	\dot G^{1100}&=& 4\eta(\alpha)\frac{\partial}{\partial x
	}\left(g_{1}(x,y)H_{IX}^{-1}\right)G^{2000}- \left(H_{IX}^{-1}-p_{x}^{2}H_{IX}^{-3}\right)  G^{0200} \nonumber \\ 
	\dot G^{0200}&=& 8\eta(\alpha)\frac{\partial}{\partial x
	}\left(g_{1}(x,y)H_{IX}^{-1}\right) G^{1100} - \frac{8p_{x}}{H_{IX}^{3}}\eta(\alpha) g_{1}(x,y) G^{0200}\nonumber \\
	\dot G^{0020}&=& \frac{8\sqrt{3}p_{y}}{H_{IX}^{3}}\eta(\alpha) g_{2}(x,y) G^{0020} -2\left(H_{IX}^{-1}-p_{y}^{2}H_{IX}^{-3}\right) G^{0011}\nonumber \\ 	
	\dot G^{0011}&=&   4\sqrt{3}\eta(\alpha)\frac{\partial}{\partial y
	}\left(g_{2}(x,y)H_{IX}^{-1}\right) G^{0020} -  \left(H_{IX}^{-1}-p_{y}^{2}H_{IX}^{-3}\right) G^{0002} \nonumber \\
	\dot G^{0002}&=& 8\sqrt{3}\eta(\alpha)\frac{\partial}{\partial y
	}\left(g_{2}(x,y)H_{IX}^{-1}\right)G^{0011} - \frac{8\sqrt{3}p_{y}}{H_{IX}^{3}}\eta(\alpha)g_{2}(x,y) G^{0002}. 
\end{eqnarray}

These are two sets of coupled and highly non-linear equations providing the evolution of the quantum modified anisotropies, we employ a numerical method to obtain the dynamics of this system, and its trajectories. 

To obtain the dynamics of classical variables ($q$, ${\beta'}_{+}$, ${\beta'}_{-}$) and their corresponding momenta, we use the Hamiltonian (\ref{TrHam}). To this end we construct $H_{Q}$ using equation (\ref{generalHq}) with three degrees of freedom. 

Up to second order in momenta we obtain the following
\begin{eqnarray}\label{3p2order}
   H_{Q}&=& \mathcal{K}- \frac{9}{4}A + D + G + \frac{1}{2}\frac{\partial^{2}\mathcal{K}}{\partial q^{2}}B+ \frac{1}{2}\frac{\partial^{2}\mathcal{K}}{\partial {\beta'}_{+}^{2}}E+\frac{1}{2}\frac{\partial^{2}\mathcal{K}}{\partial {\beta'}_{-}^{2}}H,   
\end{eqnarray}
where the momenta for three degrees of freedom is defined as follows
\begin{eqnarray}\label{G3par}
	G^{a b c d e f} &:=& \langle  (\hat p-\langle \hat p\rangle )^{a}(\hat q-\langle \hat q\rangle)^{b} (\hat p_{+}-\langle \hat p_{+}\rangle )^{c}(\hat {\beta'}_{+}-\langle \hat {\beta'}_{+}\rangle)^{d} (\hat p_{-}-\langle \hat p_{-}\rangle )^{e}(\hat {\beta'}_{-}-\langle \hat {\beta'}_{-}\rangle)^{f}\rangle_{\text{\text{Weyl}}}.
\end{eqnarray}

In the table (\ref{table:1}), the second order momenta are summarized
\begin{table}[!h] 
\begin{center}
\begin{tabular}{|c |c |c |} 
 \hline
 $A= G^{20 00 00} $ & $D= G^{00 20 00}$ & $G= G^{00 00 20}$ \\ 
 \hline
 $B= G^{02 00 00} $ & $E= G^{00 02 00}$ & $H= G^{00 00 02}$ \\
 \hline
 $C= G^{11 00 00} $ & $F= G^{00 11 00}$ &$J= G^{00 00 11}$ \\
 \hline
\end{tabular}
\caption{ Second order momenta for three degrees of freedom.}
\label{table:1}
\end{center}
\end{table}

The equations of motion obtained from (\ref{3p2order}) form a system of fifteen coupled equations that describes the dynamics of the Mixmaster model. Equations for the classical variables are 
\begin{eqnarray}\label{eM3D2D}
   \dot p &=& -\frac{3 (4 \pi)^4}{8 \pi}\frac{\partial}{\partial q}\left(q^{2/3}V(q,{\beta'}_{+},{\beta'}_{-})\right)- \frac{1}{2}\frac{\partial^{3}\mathcal{K}}{\partial q^{3}}B -\frac{1}{2}\frac{\partial^{3}\mathcal{K}}{\partial q\partial {\beta'}_{+}^{2}}E-\frac{1}{2}\frac{\partial^{3}\mathcal{K}}{\partial q\partial {\beta'}_{-}^{2}}H,\nonumber \\
   \dot {p'}_{+} &=& -\frac{3 (4 \pi)^4 q^{2/3}}{8 \pi}\frac{\partial}{\partial {\beta'}_{+}}V(q,{\beta'}_{+},{\beta'}_{-})- \frac{1}{2}\frac{\partial^{3}\mathcal{K}}{\partial {\beta'}_{+} \partial q^{2}}B -\frac{1}{2}\frac{\partial^{3}\mathcal{K}}{\partial {\beta'}_{+}^{3}}E-\frac{1}{2}\frac{\partial^{3}\mathcal{K}}{\partial {\beta'}_{+} \partial {\beta'}_{-}^{2}}H ,\nonumber \\
   \dot {p'}_{-} &=& -\frac{3 (4 \pi)^4 q^{2/3}}{8 \pi}\frac{\partial}{\partial {\beta'}_{-}}V(q,{\beta'}_{+},{\beta'}_{-})- \frac{1}{2}\frac{\partial^{3}\mathcal{K}}{\partial {\beta'}_{-} \partial q^{2}}B -\frac{1}{2}\frac{\partial^{3}\mathcal{K}}{\partial {\beta'}_{-}\partial {\beta'}_{+}^{2}}E-\frac{1}{2}\frac{\partial^{3}\mathcal{K}}{\partial {\beta'}_{-}^{3}}H ,\nonumber \\
    \dot q &=& -\frac{9}{2}p ,\nonumber \\
   \dot {\beta'}_{+} &=& 2{p'}_{+} ,\nonumber \\
   \dot {\beta'}_{-} &=& 2{p'}_{-} ,
\end{eqnarray}
while for quantum variables we have
\begin{eqnarray}\label{QeM3D2D}
   \dot A &=& -2 \frac{\partial^{2}\mathcal{K}}{\partial q^{2}}C, \hspace{1.5cm} \dot D = -2 \frac{\partial^{2}\mathcal{K}}{\partial {\beta'}_{+}^{2}} F, \hspace{1.5cm} \dot G = -2 \frac{\partial^{2}\mathcal{K}}{\partial {\beta'}_{-}^{2}} J, \nonumber \\
   \dot B &=& -9C ,\hspace{2.2cm} \dot E =  4F, \hspace{2.6cm}\dot H = 4J ,\nonumber \\
   \dot C &=&  -\frac{9}{2}A -\frac{\partial^{2}\mathcal{K}}{\partial q^{2}}B, \hspace{0.7cm}\dot F =  2D- \frac{\partial^{2}\mathcal{K}}{\partial {\beta'}_{+}^{2}} E,\hspace{1cm} \dot J = 2G- \frac{\partial^{2}\mathcal{K}}{\partial {\beta'}_{-}^{2}} H. 
\end{eqnarray} 

\subsection{Canonical effective dynamics}

Using equations (\ref{TrHam}) and (\ref{vall2pairs}), we construct the effective Hamiltonian $\mathcal{K}_{all}$ defined as
\begin{eqnarray}\label{HePOT}
   \mathcal{K}_{\text{all}} &=& -\frac{9}{4}(p^{2}+p_{1}^{2}) + ({p'}_{+}^{2}+p_{2}^{2})+({p'}_{-}^{2}+p_{3}^{2})+ V_{\text{eff}}, 
\end{eqnarray}
where
\begin{eqnarray}\label{Vpmef}
   V_{\text{eff}}&=&\left(\frac{s_{2}^{2}s_{3}^{2}+s_{1}^{2}s_{3}^{2}+s_{1}^{2}s_{2}^{2}}{8s_{1}^{2}s_{2}^{2}s_{3}^{2}}\right)+\frac{3(4 \pi)^{4}}{8k} \left[(q+s_{1})^{2/3} V_{\mathcal{H}}^{+}+(q-s_{1})^{2/3}V_{\mathcal{H}}^{-}\right],
   \end{eqnarray}
and
\begin{eqnarray}\label{Vpme}
   V_{\mathcal{H}}^{\pm} &=&e^{-\frac{8({\beta'}_{+}\pm s_{2})}{(q \pm s_{1})}}- 4e^{-\frac{2({\beta'}_{+} \pm s_{2})}{(q+s_{1})}}\cosh(\frac{2\sqrt{3}({\beta'}_{-}\pm s_{3})}{(q\pm s_{1})}) 
	+ 2e^{\frac{4({\beta'}_{+}\pm s_{2})}{(q\pm s_{1})}}\left[\cosh(\frac{4\sqrt{3}({\beta'}_{-}\pm s_{3})}{(q\pm s_{1})})-1\right].
   \end{eqnarray}
   
In this scheme, the new pairs of canonical variables ($s_{i}$, $p_{s_{i}}$) encode all the quantum information of the system. The equations of motion obtained from  this effective Hamiltonian are 
\begin{eqnarray}\label{EpAni}
	\dot q&=&  -\frac{9}{2}p, \hspace{1cm} \dot p = -\frac{1}{8}\frac{\partial}{\partial q} (V_{\mathcal{H}}^{+}+V_{\mathcal{H}}^{-})    ,\nonumber  \\
    \dot {\beta'}_{+}&=&  2{p'}_{+},  \hspace{1cm} \dot {p'}_{+} =  -\frac{1}{8}\frac{\partial}{\partial {\beta'}_{+}} (V_{\mathcal{H}}^{+}+V_{\mathcal{H}}^{-}), \nonumber  \\
    \dot {\beta'}_{-}&=&  2{p'}_{-}, \hspace{1cm} \dot {p'}_{-} =  -\frac{1}{8}\frac{\partial}{\partial {\beta'}_{-}} (V_{\mathcal{H}}^{+}+V_{\mathcal{H}}^{-}),\nonumber  \\
    \dot s_{1}&=&  -\frac{9}{2}p_{1}, \hspace{1cm} \dot p_{1}= \frac{1}{4 s_{1}^{3}}-\frac{1}{8}\frac{\partial}{\partial s_{1}} (V_{\mathcal{H}}^{+}+V_{\mathcal{H}}^{-}),\nonumber  \\
    \dot s_{2}&=&  2p_{2}, \hspace{1cm} \dot p_{2}= \frac{1}{4 s_{2}^{3}} -\frac{1}{8}\frac{\partial}{\partial s_{2}} (V_{\mathcal{H}}^{+}+V_{\mathcal{H}}^{-}),\nonumber  \\
    \dot s_{3}&=&  2p_{3}, \hspace{1cm} \dot p_{3}= \frac{1}{4 s_{3}^{3}} -\frac{1}{8}\frac{\partial}{\partial s_{3}} (V_{\mathcal{H}}^{+}+V_{\mathcal{H}}^{-}). 
\end{eqnarray}

\section{Numerical solution.}

For one degree of freedom, the momenta $G^{a,b}$ are defined as
\begin{eqnarray}
 G^{a,b} &:=& \langle  (\hat p-p )^{a} (\hat q-  q)^{b} \rangle_{\text{Weyl}} \nonumber
\end{eqnarray}
where $p:=\langle \hat p \rangle,$ and $ q:=\langle \hat q \rangle$.
Using the Gaussian wave function (\ref{GaussP}), the initial conditions for the momenta $G^{a,b}$ can be determined. For instance, the initial condition for $G^{2,0}$ is  
\begin{eqnarray}
G^{2,0} &=& \langle \psi_{\sigma}^{*}(\hat q)|(\hat p-  p )^{2}|\psi_{\sigma}(\hat q)\rangle \nonumber \\
&=& \eta \int_{-\infty}^{\infty} e^{\frac{-(\hat q-\langle \hat q \rangle)^{2}}{\sigma^{2}}}\left(\zeta-2\langle \hat q \rangle +\hat q^{2}\right) d\hat q \nonumber \\
&=&\frac{\hbar ^{2}}{2 \sigma ^{2}}, 
\end{eqnarray}
where $\eta =\frac{-\hbar^2}{\sqrt{\pi}\sigma^{5}}$, and $\zeta= \langle \hat q\rangle ^{2}-\sigma^{2}$ are constants. Following a similar procedure, the second order initial values for momenta are 
\begin{eqnarray}\label{InicialConditonsIX2order}
	G^{2,0}&=&\frac{\hbar ^{2}}{2 \sigma ^{2}}, \nonumber \\ 
	G^{1,1}&=&0,  \nonumber \\  
    G^{0,2}&=&\frac{ \sigma ^{2}}{2} .
\end{eqnarray}

On the other hand, for two degrees of freedom the momenta $G^{a,b,c,d}$ are defined as
\begin{eqnarray}
 G^{a,b,c,d} &:=& \langle  (\hat p_{1}-  p_{1} )^{a} (\hat q_{1}-  q_{1} )^{b} \nonumber \\
 &\times&(\hat p_{2}-  p_{2} )^{c}(\hat q_{2}- q_{2} )^{d} \rangle_{\text{Weyl}}, \nonumber
\end{eqnarray}
where $p_{1}:=\langle \hat p_{1} \rangle, q_{1}:=\langle \hat q_{1} \rangle, p_{2}:=\langle\hat p_{2} \rangle,$ and $ q_{2}:=\langle\hat p_{2} \rangle$. Employing a similar procedure for the one degree of freedom case, the initial conditions for the second order moments, with two degree of freedom $G^{abcd}$ are
\begin{eqnarray}\label{InicialConditons2o3p}
	G^{2000}&=&G^{0020} =\frac{\hbar ^{2}}{2 \sigma ^{2}}, \nonumber \\ 
	G^{1100}&=&G^{0011} =0,  \nonumber \\  G^{0200}&=&G^{0002} =\frac{ \sigma ^{2}}{2} .
\end{eqnarray}

For three degrees of freedom we get
\begin{eqnarray}\label{InicialConditon2order3P}
G^{200000}&=&G^{002000}=G^{000020} =\frac{\hbar ^{2}}{2 \sigma ^{2}}, \nonumber \\ 
	G^{110000}&=&G^{001100}= G^{000011}=0,  \nonumber \\  G^{020000}&=&G^{000200}= G^{000002}=\frac{ \sigma ^{2}}{2} .
\end{eqnarray}

The set of equations (\ref{EFEDynamicsiX}) and (\ref{appendix1}) describes the dynamics of anisotropies and quantum momenta of the Mixmaster model. The coupling between both kinds of variables generates a quantum backreaction that modifies the classical evolution of the system, which we explore now.

\begin{figure}[ht]
        \centering  
   	\begin{subfigure}[b]{0.45\textwidth}
   		\centering   \includegraphics[width=\textwidth]{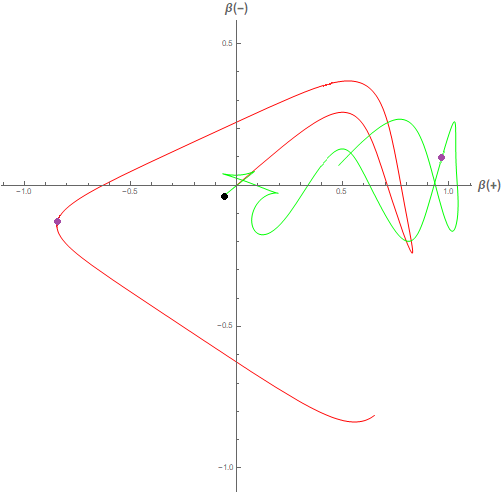}
   		\caption{Classical (red) and effective evolution (green) for anisotropies up to $t=6$.}
   		\label{BetaSingle1}
   	\end{subfigure}
   	\hfill
   	\begin{subfigure}[b]{0.45\textwidth}
   		\centering
\includegraphics[width=\textwidth]{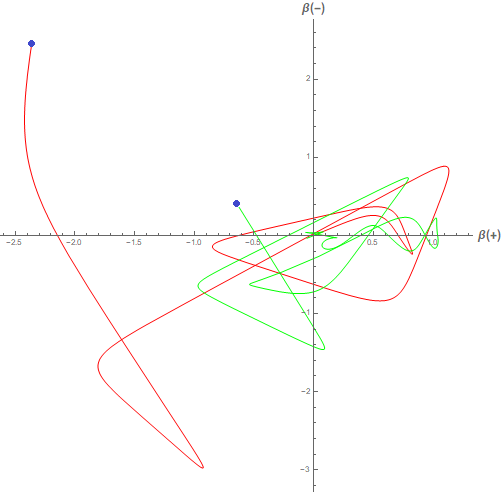}   		\caption{Classical (red) and effective evolution (green) for the anisotropies up to $t=20$.}
   \label{BetaSingle2}
   	\end{subfigure}
   	\caption{Comparison between classical (red) and effective evolution (green) for the anisotropies of the Mixmaster model. The initial singularity corresponds to $\abs{\beta_{\pm}} \rightarrow \infty $. The dots represent the position at equal times. The initial conditions are $\beta_{+}(0)=0$, $\beta_{-}=0$, $p_{+}=100$, and $p_{-}=64$, and those of eq.(\ref{InicialConditonsIX2order}) for the momenta with $\sigma =2.2$.}
    \label{BetaSingle}
   \end{figure}

We employ initial conditions (\ref{InicialConditons2o3p}) to generate, numerically, the evolution of the system. In figure \ref{BetaSingle} we show the modification of the classical trajectories (red) due the quantum effects. In figure \ref{BetaSingle1} the black dot denotes the starting point of both trajectories classical and effective (green), while the purple dots correspond to both trajectories at the same time. However, unlike the straight line behavior of classical trajectories near the initial singularity as $\abs{\beta_{\pm}} \rightarrow \infty $, as discussed in section \ref{seccion2}, and shown in figure \ref{1trayectoryClassical}, the interaction of the quantum momenta with the classical system increases the changes in direction of the trajectory for the semiclassical particle at early times, drastically changing its linear behavior. Figure \ref{BetaSingle2} represents the same trajectories for different evolution times.
   
\begin{figure}[ht]
   	\centering
   	\begin{subfigure}[b]{0.3\textwidth}
   		\centering
\includegraphics[width=\textwidth]{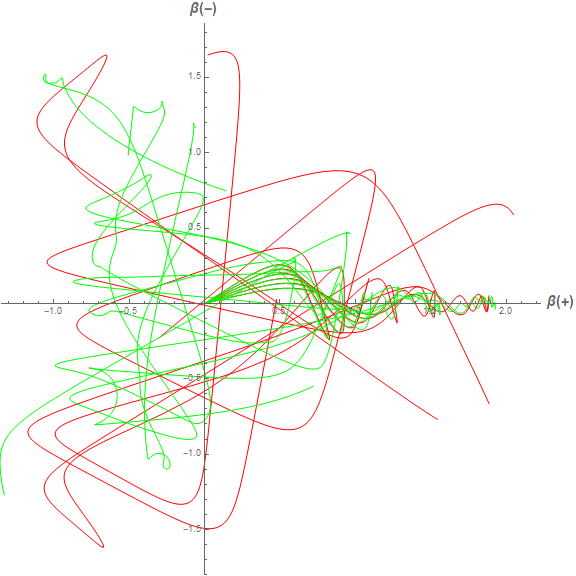}
   	\caption{Classical and quantum evolution for $\sigma =0.08$.}
   		\label{AniCH1}
   	\end{subfigure}
   	\hfill
   	\begin{subfigure}[b]{0.3\textwidth}
   		\centering   		\includegraphics[width=\textwidth]{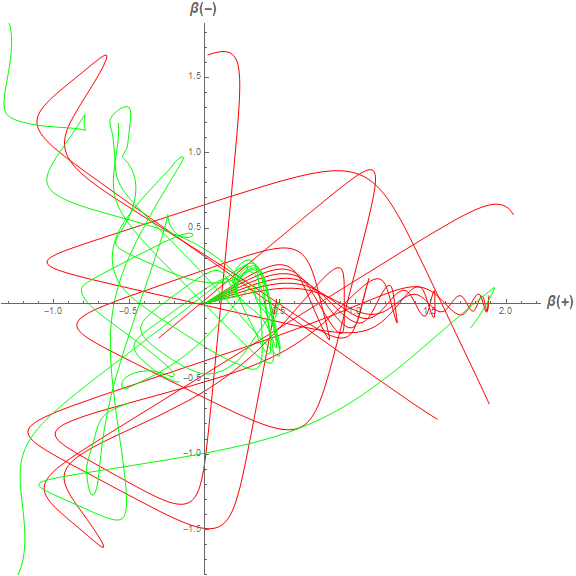}
   		\caption{Classical and quantum evolution for $\sigma =0.6$.}
   		\label{AniCH2}
   	\end{subfigure}
    \hfill
   	\begin{subfigure}[b]{0.3\textwidth}
   		\centering   \includegraphics[width=\textwidth]{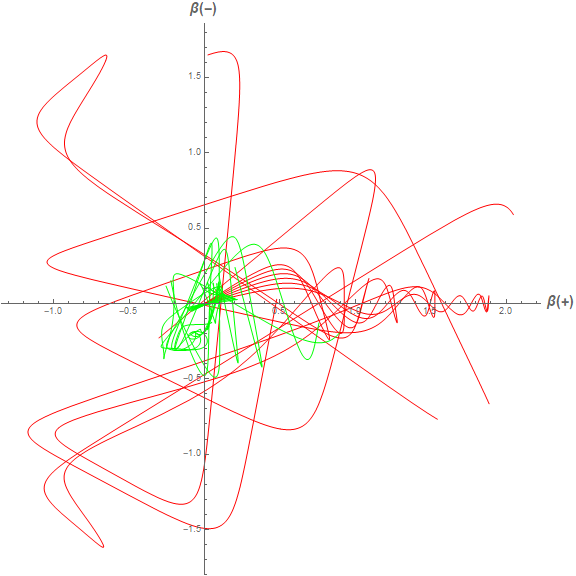}
   	\caption{Classical and quantum evolution for $\sigma =2.6$.}
   		\label{AniCH3}
   	\end{subfigure}
   	\caption{Comparison between classical (Red) and effective evolution (Green) for the anisotropies of the Mixmaster model. The initial conditions for the classical model are $\beta_{+}(0)=0$, $\beta_{-}=0$, $p_{+}=100$, and $p_{-}=8i$ (where $i=1,2,3 \cdots ,10$ corresponds to a different trajectory). The initial conditions for the effective evolution are $\beta_{+}(0)=0$, $\beta_{-}=0$, $p_{+}=100$, $p_{-}=8i$ and those of eq.(\ref{InicialConditonsIX2order}) for the momenta.}
   	\label{AniCH}
   \end{figure}

In figure \ref{AniCH} we show a comparison between several classical and semiclassical trajectories, displaying chaotic behavior \cite{bojowald2023chaotic}.

Additionally in Figure (\ref{EffRemo2}) we show the classical and semiclassical evolution in the phase space diagram ($ p ,Log(q)$), where $q$ is related to the scale factor through $q=\sqrt{a}$. In this diagram $t=0$ occurs when $Log(q) \rightarrow -\infty$, that is,  $a\rightarrow 0$. The classical trajectory (blue) displays $Log(q)  \rightarrow -\infty $, that is, it contains the initial singularity, while the effective trajectories in momenta (black) and canonical potential (purple)  $\abs{Log(q)} < \infty $, thus removing the initial singularity.
\begin{Figure}
	\centering
\captionsetup{type=figure}
\includegraphics[width=0.9\linewidth]{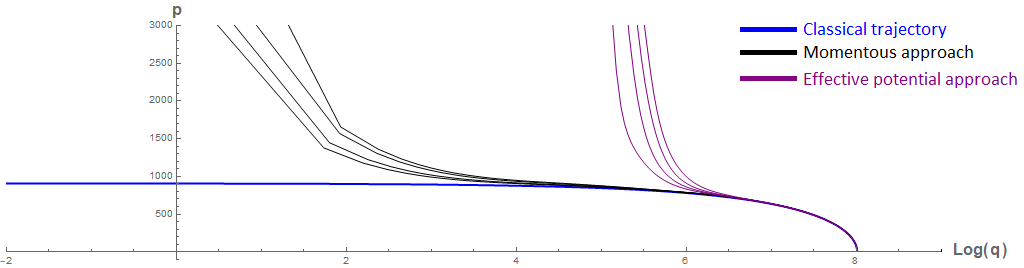}
	\captionof{figure}{Phase space diagram of $ Log(q) $ and $p$. Unlike the classical evolution (blue), which contains the initial singularity at t=0, the effective trajectories avoid it  because $\abs{Log(q)} < \infty $. The different trajectories corresponds to a different values of  $\sigma$. The initial conditions are $q(0)=0.1$, $p(0)=1200$, and $\sigma=6, \sigma=8, \sigma=10$, and $ \sigma=12$.}
  \label{EffRemo2}
 \end{Figure}

In figure \ref{EffRemo3} we show classical and  semiclassical trajectories within the all order potential method for different initial conditions of quantum variables $s_{i}$.  
 \begin{Figure}
	\centering
\captionsetup{type=figure}
\includegraphics[width=0.9\linewidth]{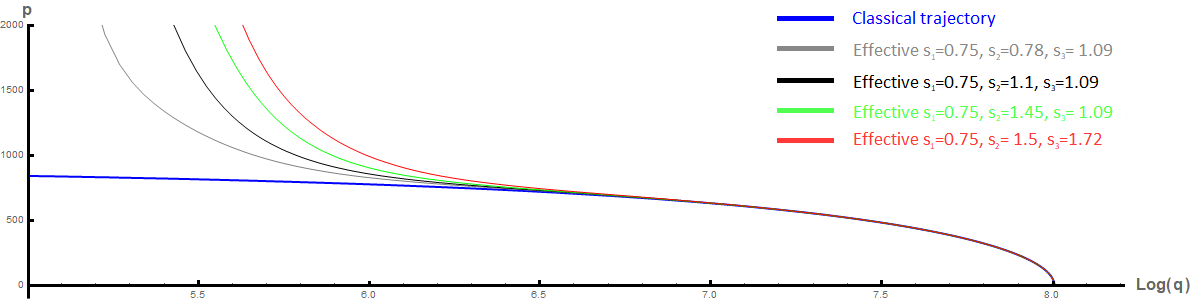}
	\captionof{figure}{Comparison between the classical evolution (blue) and semiclassical with effective potential in the phase space diagram $Log(q) $ and $ p$. For the classical trajectory $Log(q) \rightarrow -\infty$, while in the semiclassical ones $\abs{Log(q)} < \infty $, removing the initial singularity. The initial conditions are $q(0)=0.1$, $p(0)=1200$, and  $\sigma=6, \sigma=8, \sigma=10, \sigma=12$.}
  \label{EffRemo3}
 \end{Figure}

\section{Conclusions}

In this article the quantum Mixmaster  model was analyzed in a semiclassical setting, finding important differences with respect to the classical model. In particular, by studying semiclassical trajectories describing the evolution of anisotropies we show that the initial singularity is avoided as a result of quantum back reaction.

In the classical Mixmaster model, the universe is subject to a time dependent potential evolving in a space of anisotropies, and has a singularity at $t=0$, where the interaction with the potential walls is less frequent. 
We apply a canonical transformation in the classical Hamiltonian to obtain an expression more suitable to implement the effective analysis for the quantum model. Under this formulation it is possible to make a direct comparison between the quantum corrected behavior and that of the classical model by means of effective trajectories.

The behavior displayed by the quantum model derived from the effective Hamiltonian (\ref{eHM}) shows that, although the classical evolution is drastically modified due to quantum backreaction through quantum momenta, it retains its chaotic behaviour regardless of initial conditions taken into account, in agreement with recent similar studies \cite{bojowald2023chaotic}. 

In the classical evolution, the initial singularity is reached as $\abs{\beta_{\pm}} \rightarrow \infty $. Because the trajectory is a straight line in that case, it means that the particle does not interact with the classical potential near $t=0$. In the quantum regime, the interaction of the particle with the potential is stronger due to back reaction, generating more dispersions, and preventing the particle from reaching $\abs{\beta_{\pm}} \rightarrow \infty $, thus avoiding the singularity.

After performing a canonical transformation we obtain an effective Hamiltonian (\ref{3p2order}) in terms of anisotropies, the scale factor $a$ and their quantum momenta. Unlike the classical behaviour, the effective evolution shows that quantum variables impose a minimum lower bound on the volume of the universe, as shown in figure \ref{EffRemo2}, i.e, the initial singularity is removed. 

Finally, the evolution obtained from the canonical Hamiltonian (\ref{HePOT}) also displays a singularity avoidance (Figure \ref{EffRemo3}). The Hamiltonian (\ref{3p2order}) is used to perform an effective analysis based in momenta up to second order $G^{a b c d}$ (Figure \ref{EffRemo2}), while the Hamiltonian (\ref{HePOT}) is employed to generate the effective dynamics of the system in terms of canonical variables ($s_{i}, p_{s_{i}}$) and effective potential.

The main difference between the momenta and the canonical potential descriptions is the need to truncate the dynamical system for the former. This in turn derives in that the minimum value for $a$ is different in each case, being larger when the canonical potential is used. For the present analysis both descriptions were possible and, as shown, both give similar results; however, there are systems for which one description cannot be applied, but the other is.

In our analysis we have applied a canonical transformation that renders the effective Hamiltonian (\ref{HIx}) explicitly as kinetic plus potential terms. This transformation  allows the application of the generalized potential method, and with this obtain the evolution of the scale factor and the anisotropies of the system from the same system of equations, which is not possible in the $G^{a,b}$ formulation. There exist similar effective analysis of the Mixmaster model focusing on the study of anisotropies \cite{brizuela2022semiclassical, bojowald2023chaotic, de2023mixmaster}. 

Our study can be used to generalize the analysis of anisotropic cosmological models into inhomogeneous models, where we expect to obtain interesting  results in this effective description \cite{bojowald2021canonical, bojowald2022quasiclassical, brizuela2019moment}.

\addcontentsline{toc}{chapter}{Bibliografía}
\nocite{*} 
\bibliographystyle{unsrt}
\bibliography{References}

\end{document}